\documentclass[aps,prd,twocolumn,showpacs,floats,floatfix,letterpaper,nofootinbib,superscriptaddress,]{revtex4-2}

\usepackage{hyperref}
\usepackage{amssymb,amsmath,latexsym,mathrsfs}
\usepackage{graphicx,subfigure}
\usepackage{epsfig}
\usepackage{varioref,xr-hyper}
\usepackage{color}
\usepackage{multirow}
\usepackage{array}
\hypersetup{
	colorlinks   = true,
	citecolor    = blue
}

\begin{document}

\title{Cosmological parameter constraints using phenomenological symbolic expressions: On the significance of symbolic expression complexity and accuracy}

\author{S. M. Koksbang} 
\email{koksbang@cp3.sdu.dk}
\affiliation{CP3-Origins, University of Southern Denmark, Campusvej 55, DK-5230 Odense M, Denmark}

\begin{abstract}
Phenomenological models are widely used in cosmology in relation to constraining different cosmological models, with two common examples being cosmographic expansions and modeling the equation-of-state parameter of dark energy. This work presents a study of how using different phenomenological expressions for observables and physical quantities versus using physically motivated, derived expressions affects cosmological parameter constraints. The study includes the redshift-distance relation and Hubble parameter as observables, and the dark energy equation-of-state parameter as a physical quantity, and focuses on constraining the cosmological parameter $\Omega_{\Lambda}$. The observables and equation-of-state parameter are all modeled both using the physical, derived expressions and a variety of phenomenological models with different levels of accuracy and complexity. The results suggest that the complexity of phenomenological expressions only has minor impact on the parameter constraints unless the complexity is very high. The results also indicate that statistically significantly different results can be expected from parameter constraints using different phenomenological models if the models do not have {\em very} similar accuracy. This suggests that a good practice is to use multiple phenomenological models when possible, in order to assess the model dependence of results. Straightforward examples of this is that results obtained using cosmographic expansions should always be checked against similar results obtained with expansions of other order, and when using phenomenological models such as for the equation-of-state parameters, robustness of results could be assessed using fitted models from symbolic regression, similar to what is done in this study.
\end{abstract}

\maketitle

\section{Introduction}\label{sec:intro}
Using data to constrain cosmological model parameters such as density parameters, curvature and the Hubble constant is one of the main cornerstones of modern cosmology. Simply put, with this approach cosmologists seek to estimate values of model parameters by comparing theoretical predictions/expressions for observables with observations and tuning the model parameters until maximal agreement between observations and theoretical prediction is achieved. Parameter constrains are usually achieved by assuming that the Universe behaves as a Friedmann-Lemaitre-Robertson-Walker (FLRW) universe on large scales. In this case, symbolic expressions for observational relations such as the redshift-distance relation are often straightforward to {\em derive} mathematically.
\newline\indent
In recent years, cosmological parameter inference has been pushed toward methods for inferring parameter constraints which rely less on specific cosmological models. A popular method in this vain is to do a cosmographic expansion which is basically a Taylor expansion of the observable. The redshift or a reparametrization of this is typically used as the expansion parameter. Clearly, a cosmographic expansion is not model-independent since it is usually based on Taylor expanding the observable (mostly the redshift-distance relation) as it is given in the FLRW model. The goal with the approach is then to constrain standard cosmological parameters such as the Hubble constant and deceleration parameter without making an assumption regarding e.g. the sign of the spatial curvature or the form of the equation-of-state parameter of dark energy. In addition to clearly still being model dependent, results obtained using cosmographic expansions will depend on the choice of expansion parameter and truncation order -- the latter is not only true for a Taylor expansion but also for other similar expansions such as Principle Component Analysis (see e.g. \cite{PCA_1, PCA_2} for cosmology related examples). The former is e.g. illustrated in \cite{issue_1} while the latter is discussed thoroughly in \cite{issue_2}. A main result in \cite{issue_2} is the illustration that using different (phenomenological) polynomial representations for observational relations can lead to significantly different parameter inferences. This is an important point considering that the cosmological literature contains several examples of using phenomenological representations for observables, including e.g. polynomials and splines \cite{spline_1, spline_2, spline_3} and linear combinations of pre-defined ``basis functions'' \cite{basis}, in addition to work based on cosmographic expansions such as \cite{cosmograph_1, cosmograph_2, cosmograph_3, cosmograph_4, cosmograph_5, cosmograph_6, cosmograph_7, cosmograph_8, cosmograph_9, cosmograph_10}. Even for the increasingly popular Gaussian Process which has been used for e.g. constraining the Hubble constant \cite{GP_Hubble}, curvature \cite{GP_curvature} and matter density \cite{GP_matter} and for reconstructing the dark energy equation-of-state parameter \cite{GP_reconstruct}, there are similar possible issues. For instance, in \cite{GP_issue_1} it was pointed out that the results from a Gaussian Process depends on the choice of hyper parameters and kernels. The work of \cite{GP_issue_1} therefore focuses on studying how different choices of kernels and hyper parameters affect model constraints. Similarly, \cite{GP_issue_2} compares parameter constraints using the Gaussian Process with different kernels as well as different FLRW models. Although the authors find agreement between the methods within 1 standard deviation, the actual values of the best fit Hubble constant found by the different methods disagree by as much as almost 10\%. See e.g. \cite{colgain_1, colgain_2, colgain_3} for other discussions on the model dependence of cosmographic expansions and Gaussian Processes.
\newline\indent
It is not only when seeking model-independent parameter constraints that phenomenological models can be useful. Phenomenological models are for instance routinely used when seeking to learn about dark energy where a variety of different phenomenological descriptions for the dark energy equation-of-state parameter are used in the literature (see e.g. \cite{PCA_2, omegaDE_1, omegaDE_2, omegaDE_3, omegaDE_4, omegaDE_5, omegaDE_6, omegaDE_7, omegaDE_8, omegaDE_9, omegaDE_10, omegaDE_11, omegaDE_12, omegaDE_13, omegaDE_14, omegaDE_15}). Also the deceleration parameter is often considered using phenomenological models (see e.g. \cite{omegaDE_8, q1, q2, q3, q4, q5, q6, q7, q8, q9, q10, q11} for examples). In the case of dark energy, the use of phenomenological models is necessary because the true equation-of-state parameter is unknown. Similar situations arise regarding cosmic backreaction \cite{fluid1, fluid2, bc_review1, bc_review2, bc_review3} as well as the mean redshift drift\footnote{Redshift drift is the temporal evolution of the redshift of an astrophysical source. It was first discussed in \cite{dz_sandage, dz_Mcvittie} where it was considered with the FLRW models. As was first realized in \cite{another_look}, in inhomogeneous cosmological models, the mean redshift drift is not in general equal to the drift of the mean redshift and can therefore not be described by a simple analog to the FLRW case.}. This was discussed in \cite{inhomo_AIFeynman1, inhomo_AIFeynman2} where phenomenological models of cosmic backreaction and the mean redshift drift were obtained using the publicly available machine learning algorithm AI Feynman \cite{AIFeynman1, AIFeynman2} which performs symbolic regression. For physics problems, AI Feynman tends to outperform algorithms based on genetic algorithms which is a popular strategy for symbolic regression, used in e.g. \cite{genetic_SN} with the focus on analyzing supernova data, \cite{genetic_GL} studying gravitational lens inversion, \cite{genetic_GW} studying gravitational waves, \cite{genetic_tests} analyzing data in relation to a cosmological consistency test, and \cite{genetic_gaussian_cosmo_compare} which compares model constraints obtained using Gaussian Processes, gplearn's genetic algorithm\footnote{https://gplearn.readthedocs.io/en/stable/} and cosmographic expansions.
\newline\newline
The examples above illustrate that phenomenological models are often used in cosmology when seeking to constrain cosmological parameters and that while different models will lead to different constraints, there are a variety of different phenomenological models to choose from, without any clear \emph{a priori} way to prioritize between them. Especially looking into a future where phenomenological models obtained from machine learning such as symbolic regression or Gaussian Processes will inevitably become more common, it seems prudent to study further what the consequences are for parameter extraction when different phenomenological versus true/theoretically derived models are used. This is the focus of the work presented here, which consists of a comparison of parameter constraints obtained using different phenomenological symbolic expressions for the redshift-distance and Hubble parameter in standard FLRW models. In addition, parameter constraints will be obtained using different phenomenological models for the equation-of-state parameter of dark energy for a specific dark energy model. In order to compare the phenomenological models, these will be characterized through complexity and accuracy measures. In section \ref{sec:observables} the phenomenological expressions for the two observable relations and the equation-of-state parameter will be presented and compared using measures of accuracy and complexity. Section \ref{sec:results} then compares the parameter constraints obtained using the different phenomenological expressions while a summary with discussion and concluding remarks is provided in section \ref{sec:summary}.

\section{Observables}\label{sec:observables}
This section serves to introduce the observables that will be used for obtaining parameter constraints in section \ref{sec:results}. Below, two different types of observables including their FLRW {\em theoretically derived} expressions will first be presented before moving on to introducing the dark energy model which will be considered. Afterwards, phenomenological models will be presented and discussed.
\\\\
The first observable that will be considered is the redshift-distance relation which can be obtained observationally from supernovae through their flux magnitude. This magnitude can be directly related to the luminosity distance \cite{JLA} which can again be related to the angular diameter distance $D_L$ through the distance-duality relation, $D_L = D_A(1+z)^2$ \cite{Etherington}. In the FLRW limit, the luminosity distance is given by
\begin{align}
	D_L = \frac{1+z}{\sqrt{K/R_0^2}}\sin\left(\sqrt{K/R_0^2}\int_0^z\frac{dz'}{H(z')} \right),
\end{align}
where $R_0$ is the spatial curvature radius and $K$ indicates the sign of the curvature. We will only be concerned with the limit of vanishing curvature ($K=0$) in which case the expression reduces to
\begin{align}\label{eq:DL}
	D_L^{\rm flat} = (1+z)\int_0^z\frac{dz'}{H(z')}.
\end{align}
\\\\
We will consider the redshift-distance relation because it is of chief importance among cosmological observables; the redshift-distance relation not only represents the basis for interpreting supernova observations cosmologically but is also important for cosmological parameter constraints obtained using Baryon Acoustic Oscillations (BAO) measurements (see e.g. \cite{BAO0, BAO1, BAO2, BAO3, BAO4} and section 4 of \cite{BAO5}) and Cosmic Microwave Background (CMB) data. The latter is e.g. exemplified by the appearance of $D_A$ in the shift parameter (see e.g. \cite{CMB1, CMB2, CMB3}) and its effect on CMB peak positions (see e.g. eq. 21a in \cite{CMB1}).
\newline\newline
It is interesting to also consider other types of observables in order to understand if the obtained results regarding the (in-)significance of complexity and accuracy of the phenomenological models is specific to the redshift-distance relation or if similar results hold for other observables. Indeed, in principle it would be prudent to consider \emph{all} cosmological observables individually. Such huge undertaking is beyond the scope of the current study, which will include only one more observable, namely direct measurements of the Hubble parameter, $H(z)$. Within the limits of FLRW models, direct measurements of the Hubble parameter can be achieved through a variety of methods including e.g. BAO measurements \cite{BAO0} (e.g. section 1.1.3), cosmic chronometers \cite{cc}, redshift drift \cite{dz_sandage, dz_Mcvittie} and the dipole of the luminosity distance \cite{dipole}. The FLRW Hubble parameter, $H(z)$, is given by
\begin{align}\label{eq:Friedmann}
	\frac{H^2(z)}{H_0^2}  = \frac{8\pi \rho(t)}{3} - \frac{K}{R_0^2a^2}+\frac{1}{3}\Lambda.
\end{align}
There are several reasons for choosing to focus on this quantity. First of all, as discussed in e.g. \cite{H_important, H_important_intro}, direct measurements of the Hubble parameter are of high importance in cosmology. Second of all, in the FLRW limit, the redshift drift can be directly translated into a measurement of the Hubble parameter if the Hubble constant is known. Therefore, considering Hubble parameter data in the FLRW model amounts to using redshift drift data which is particularly interesting data for this study since the mean redshift drift in general (non-FLRW) cosmological models can currently only be described through phenomenological models (see \cite{inhomo_AIFeynman1, inhomo_AIFeynman2}).
\newline\newline
The above two observables will first be used to constrain model parameters within the standard $\Lambda$CDM model. For studying the significance of the complexity of phenomenological expressions of physical quantities, a flat matter+dark energy model will be considered. The dark energy equation-of-state parameter, $\omega_{\rm de}$, for the model is chosen to be the time dependent model
\begin{align}\label{eq:omegaDE}
\omega_{\rm de} = -\omega_0 \frac{\exp\left( \frac{z}{1+z}\right) }{1+z},
\end{align}
introduced in \cite{DEmodel}. In order to introduce a dependence on $\Omega_{\Lambda}=\Omega_{\Lambda,0}$ (the parameter to be constrained), $\omega_0$ will be given as
\begin{align}
	\omega_0  =\frac{1+\Omega_{\Lambda}}{\exp\left( \frac{\Omega_{\Lambda}}{1+\Omega_{\Lambda}}\right) } ,
\end{align}
where, for notational simplicity, we will use the notation $\Omega_{\Lambda} = \Omega_{\Lambda,0}= \Omega_{\rm de,0}$ and simply refer to $\Omega_{\Lambda}$ for all models in the following section.
\newline\indent
We will be concerned with obtaining phenomenological expressions approximating the above $\omega_{\rm de}$. These expressions, $\omega_i(z)$, will be combined with equations \ref{eq:Friedmann} and \ref{eq:DL} by noting that the expression for the density parameter of dark energy for these phenomenological equation-of-state parameters can be written as
\begin{align}
	\Omega_{\rm de, i} = \Omega_{\rm de,0}\exp\left(3\int_0^z\frac{dz'}{1+z'}(1+\omega_i(z')) \right) ,
\end{align}
where $\Omega_{\rm de,0}=\Omega_{\Lambda}$ is the present time density parameter of the dark energy component. The phenomenological models for $\omega_{\rm de}$, $\omega_i$, can then be used to constrain $\Omega_{\Lambda}$ through Hubble and redshift-distance data obtained by using the exact expression for the equation-of-state parameter in equation \ref{eq:omegaDE} to generate mock data.
\newline\newline
The data sets $(D_L, z)$ and $(H,z)$, will both be considered in the form of mock data sets with equidistant values of twenty data points in the redshift interval $z\in[0,2]$, with data points being the upper redshift values in each bin, i.e. data points are considered at $z = 0.1,0.2,....1.9,2.0$. The mock data sets are generated with errors of both 10\% and 1\%. Using two levels of inaccuracy of data (1\% and 10\%) makes it possible to assess how the accuracy of data affects the conclusions.

\subsection{Phenomenological symbolic expressions: Complexity and accuracy}
In the next subsection, phenomenological symbolic expressions that approximate the redshift-distance relation and $H(z)$ in FLRW models will be presented together with phenomenological expressions for $\omega_{\rm de}$. Before that, this subsection serves to discuss the methods by which the expressions have been obtained as well as the methods used for comparing the individual expressions.
\\\\
The symbolic expressions considered in this work have been obtained by training machine learning algorithms on FLRW data in the redshift interval $z\in[0,2]$ and with fixed $H_0 = 70$km/s/Mpc, vanishing curvature and $\Omega_{\Lambda}\in[0.1,0.95]$. The redshift interval was chosen because it represents the interval where most current supernova and Hubble parameter data is contained (and naturally, the mock data has been generated within this same redshift interval). The choice to keep $H_0$ fixed was made for simplicity: By only varying one model parameter, it becomes significantly faster to obtain a variety of phenomenological symbolic expressions with reasonable accuracy. $H_0$ was chosen as the fixed parameter because it merely represents a scaling of the data whereas $\Omega_{\Lambda}$ enters into the expressions for the redshift-distance relation and $H(z)$ in a somewhat more complicated manner and in \emph{different} manners in the two data sets which is a sensible requirement; otherwise there would presumably be little to gain from considering both data sets. The choice of considered data interval for $\Omega_{\Lambda}$ is loosely based on the intervals used for this parameter in current literature.
\newline\indent
There are several publicly available algorithms that perform symbolic regression, including e.g. the earlier mentioned gplearn and AI Feynman as well as, for instance, the recent Exhaustive Symbolic Regression algorithm \cite{ESR} (see e.g. also \cite{ESR2}). Here, the two algorithms AI Feynman and PySR \cite{PySR} will be used. Both codes were developed with (astro-)physics as a main motivation and both are publicly available in ready to use formats. As detailed in \cite{AIFeynman1, AIFeynman2,PySR}, the algorithms seek to identify ``good'' symbolic expressions by weighing both accuracy and complexity. It therefore seems natural to characterize the phenomenological models compared here using some version of these qualities. 
\newline\newline
{\bf Complexity:} Quantifying the ``complexity'' of a function can be done in several ways (see e.g. \cite{complexity} for examples) and it is not trivial to determine which method is the most useful in a given situation. In particular, it is not clear that the numbers indicating complexity of the symbolic expressions supplied by the AI Feynman and PySR algorithms are the most useful for comparing the complexities of the different phenomenological models presented here\footnote{PySR can actually be supplied with a user-defined complexity measure. For the work presented here, PySR was used with the default complexity measure since there was no reason to require the considered algorithm to use the same complexity measure as that used later to compare the resulting phenomenological expressions. It was therefore easier to simply use the default complexity measure for the algorithm and then compute the complexity defined in equation \ref{eq:complexity} afterwards.}. The measures of complexity used to compare the phenomenological expressions will therefore not be based on the measures supplied by the AI Feynman and PySR algorithms.
\newline\indent
The complexity of the symbolic expressions will be assessed following the method suggested in \cite{complexity}. The complexity measure proposed in \cite{complexity} fits well with intuition as it penalizes expressions that a human (cosmologist) would typically consider complex -- for instance (nested) trigonometric functions and logarithms -- instead of merely concentrating on the length of an expression. This way, a polynomial with, say, seven terms may result in a lower complexity that a single-term function with nested trigonometric functions.
\newline\indent
Using the definition in equation 1 of \cite{complexity}, the complexity of the phenomenological symbolic expressions obtained with AI Feynman and PySR will be computed recursively from bottom to top of the symbolic expression tree according to
\begin{widetext}
\begin{equation}\label{eq:complexity}
{\rm complexity}(n) =
\left\{
\begin{aligned}
&1 \,\,\,\, &&\text{if} \,\,\,\, n=\rm constant \\
&2 \,\,\,\,&&\text{if} \,\,\,\, n= \rm variable \\
&\sum{\rm complexity}(c)\,\,\,\,&&\text{if} \,\,\,\,n\in(+,-)\\
&\prod {\rm complexity}(c)\,\,\,\,&&\text{if} \,\,\,\,n\in(*,/)\\
&{\rm complexity}(c)^3 \,\,\,\,&&\text{if} \,\,\,\, n =\rm square \, root\\
&2^{{\rm complexity}(c)}\,\,\,\,&&\text{if} \,\,\,\,n\in(\sin, \cos, \tan, \exp, \log)
\end{aligned}
\right.,
\end{equation}
\end{widetext}
where complexity($n$) is the complexity of the $n$'th node and complexity(c) represents the complexity of the direct child nodes of the $n$'th node. This complexity definition will here be extended by the bottom line in the definition above also including inverse trigonometric functions.
\newline\newline
An example of making the symbolic expression tree and applying the above algorithm to compute the complexity is given in appendix \ref{app:tree_complexity}.
\newline\newline
In addition to the above defined complexity, the phenomenological expressions will also be compared with respect to their numbers of terms since this number is sometimes used as a simple measure of complexity. Note that some of the expressions presented in the following contain nested functions. To have a consistent scheme for computing the number of terms, all $\pm$ signs will be considered on equal footing. The measure of how many terms an expression contains will therefore be computed as the total number of $\pm$ appearing in the expressions. Minus signs will not be included if they could have been avoided by swapping the place of the term with another term with positive sign, i.e. the expression $-4z+2\Omega_{\Lambda} = 2\Omega_{\Lambda} -4z$ is only considered having one $\pm$ sign.
\newline\indent
Looking at standard Markov Chain Monte Carlo algorithms such as the Metropolis-Hastings algorithm \cite{HOGG} or the slight modification utilized later in this work \cite{emcee}, it does not appear as though the complexity of a function should be of significance for parameter determinations. The complexity of a function is nonetheless a simple way to compare phenomenological expressions. Additionally, it is possible that subtle computational relations lead MCMC algorithms to perform better on, say, expressions with low complexity. For instance, some functional forms may lead to {\em Nan} at specific points inside the permitted parameter space and low complexity may lead to faster function evaluation which may affect the performance of an algorithm. It therefore seems prudent to compare the phenomenological expressions in terms of their complexity.
\newline\newline
{\bf Accuracy:} It is self-evident that the accuracy of the phenomenological models is significant for obtaining model constraints. It is less clear exactly how this shows itself in relation to mean values and confidence intervals when considering phenomenological expressions with various cumulative and local inaccuracies. The accuracy of the considered phenomenological expressions will therefore be computed in several different ways. Once symbolic expressions have been obtained with the algorithms, the accuracy of the considered symbolic expressions will be computed as the mean square error of the given symbolic expression summed over the mock data points when comparing with the FLRW model used to generate the mock data. The accuracy will also be computed as the sum of mean square errors, summing over a one million equidistant grid of points contained in the data used to train the algorithms. The first accuracy score is interesting since the MCMC code that will be used for obtaining model constraints should be focusing much attention in this parameter region as long as the symbolic expressions have a fairly high accuracy. The latter accuracy score is also important since the symbolic expressions in the entire parameter and data intervals will be available for the MCMC code and hence will necessarily have some effect on the obtained results. Both these measures of (in-)accuracy are cumulative. Figures showing residuals will be used to show the local accuracy of expressions.

\subsection{Phenomenological symbolic expressions: Studied expressions}
\begin{table}
	\centering
	\begin{tabular}{|l|rccc|}
	\hline
	Model& accuracy.20 & accuracy.all & complexity & number of $\pm$\\
	\hline
	$D_1$ & 22758329 &1040598  &417 & 20 \\
	$D_2$ & 22758298 & 1040633&658 & 4\\
	$D_3$ & 22758419 & 1006715& 2209 & 19\\
	$D_4$ & 22824605 &1006715& 3113 & 8 \\
	$D_5$ & 22757455 & 1040697& $\sim 1.79\cdot 10^{103}$ & 5\\
	$D_6$ & 22758253 & 1040624 &$\sim 2^{65537}$ & 29\\
	\hline
\end{tabular}
\caption{Accuracy scores and complexity of phenomenological symbolic expressions (marked as ``models'' in the table) approximating the redshift-distance relation. The first accuracy score, ``accuracy.20'' indicates the accuracy obtained when only summing over the 20 mock data points, evaluating at the best-fit value of $\Omega_{\Lambda}$. The other accuracy score, ``accuracy.all'', was obtained as the sum of the least square errors of one million data points computed equidistantly in the intervals $z\in[0.1,2]$ and $\Omega_{\Lambda}\in[0.1,0.95]$. The lower bound on the redshift interval was chosen to be $0.1$ rather than $0$ because several expressions have poor accuracy at lower values while, at the same time, $z = 0.1$ is the lowest redshift value in the considered mock data. A significant amount of {\em Nan} were encountered when evaluating $D_3$ and $D_4$ in the million-point interval while none were encountered in the 20 point interval. To be able to make a reasonable comparison between the individual ``accuracy.all''s, this number has been divided by the number of points included in the computations. For all expressions but $D_3$ and $D_4$, this number was naturally $10^6$ while it for $D_3$ and $D_4$ was reduced to 825967 since all {\em Nan} encounters were discarded. The numbers in the column ``accuracy.20'' have not been divided by the number of data points used in the computation; this was 20 for all expressions.}
\label{table:D}
\end{table}
\begin{table}
	\centering
	\begin{tabular}{|l|rccc|}
		\hline
		Model& accuracy.20 & accuracy.all & complexity & number of $\pm$\\
		\hline
		$H_1$ & 0.00006 &4.44207 &29 &5\\
		$H_2$ & 0.00001 &0.07970 &129 & 14\\
		$H_3$ & 0.00002 & 1.23126& 213 & 6\\
		$H_4$ & 0.00069&114.58257& 514 & 2\\
		\hline
	\end{tabular}
	\caption{Accuracy scores and complexity of phenomenological symbolic expressions (marked as ``models'' in the table) approximating $H(z)$. The first accuracy score, ``accuracy.20'' indicates the accuracy obtained when only summing over the 20 mock data points, evaluating at the best-fit value of $\Omega_{\Lambda}$. The other accuracy score, ``accuracy.all'', was obtained as the sum of the least square errors of one million data points computed equidistantly in the intervals $z\in[0,2]$ and $\Omega_{\Lambda}\in[0.1,0.95]$.}
\label{table:H}
\end{table}
By generating data sets for the Hubble parameter and redshift-distance relation and providing them to AI Feynman and PySR with different hyper parameter choices, different symbolic expressions approximately describing the data have been obtained. The expressions selected for the study are presented in this section together with their accuracy and complexity values. The different expressions were selected based on wishing to obtain a variety of different phenomenological expressions with different levels of accuracy and complexity. This was possible to different degrees for the different quantities (redshift-distance data, Hubble data and equation-of-state parameter for dark energy), and by considering the results obtained by using phenomenological expressions for all three quantities collectively, the diversity in accuracy and complexity is sufficient to make (tentative) conclusions.
\\\\
The phenomenological symbolic expressions approximating the redshift-distance relation in the considered redshift and parameter interval of FLRW models obtained with AI Feynman and PySR and which will be used in the next section are\footnote{The quantities $D_i, i=1,\dots,6$ approximate the luminosity distance-redshift relation with the luminosity distance given in units of Gpc.}
\begin{widetext}
\begin{align}\label{eq:D}
	D_1 & = 0.089820z^4 - 0.634024z^3\Omega_{\Lambda} - 0.285836z^3 + 16.501933z^2\Omega_{\Lambda}^4 - 26.551723z^2\Omega_{\Lambda}^3 \\\nonumber &+ 16.517987z^2\Omega_{\Lambda}^2 - 1.939671z^2\Omega_{\Lambda} + 1.060522z^2 + 38.828078z\Omega_{\Lambda}^5 - 115.863250z\Omega_{\Lambda}^4 \\\nonumber &+ 123.598428z\Omega_{\Lambda}^3 - 58.913636z\Omega_{\Lambda}^2 + 12.668990z\Omega_{\Lambda} + 3.588830z + 45.680092\Omega_{\Lambda}^6 \\\nonumber &- 160.291754\Omega_{\Lambda}^5 + 226.591757\Omega_{\Lambda}^4 - 161.647418\Omega_{\Lambda}^3 + 60.262711\Omega_{\Lambda}^2 - 11.024120\Omega_{\Lambda} + 0.709963\\\nonumber\\\nonumber
	D_2 &= \frac{z}{10}\left( 43.292614 + \cos(z) + \left[   1.5842607\exp(\exp(\Omega_{\Lambda})) + \Omega_{\Lambda}\right]  \left[  \sin(\sin(z)) + z\right] \right)\\\nonumber\\\nonumber
	D_3 & = 0.132337z^4 - 0.620309z^3\Omega_{\Lambda} - 0.554295z^3 + 1.600080z^2\Omega_{\Lambda}^2 + 2.676984z^2\Omega_{\Lambda} \\\nonumber &+ 0.612064z^2 + 12.360111z\Omega_{\Lambda}^3 - 17.175461z\Omega_{\Lambda}^2 + 7.841053z\Omega_{\Lambda} + 2.816981z \\\nonumber &+ 7.798212\Omega_{\Lambda}^4 - 20.650463\Omega_{\Lambda}^3 + 17.691596\Omega_{\Lambda}^2 - 5.561943\Omega_{\Lambda} +0.494752\\\nonumber &+\sin^{-1}\left(-0.009464+z\cdot\left(2-\exp\left(\Omega_{\Lambda} \right) ) \right) \right)+\frac{1}{100}\sin^{-1}\left(0.21913+\cos\left(2\pi\cdot\Omega_{\Lambda} \right)  \right)\\\nonumber\\\nonumber
	D_4 &=  z\cdot \exp\left( \Omega_{\Lambda}\right) +z\cdot\exp\left( \Omega_{\Lambda}\cdot \tan^{-1}(z)+1\right)+\sin^{-1}\left( -0.009464+z\cdot\left(2-\exp\left(\Omega_{\Lambda} \right) ) \right) \right) \\\nonumber & +\frac{1}{100}\sin^{-1}\left(0.21913+\cos\left(2\pi\cdot\Omega_{\Lambda} \right)  \right)  -0.03671\\\nonumber\\\nonumber
	D_5 & =  -2.718423+\exp\left( \sqrt{z(\Omega_{\Lambda}+\pi)+1}\right) +0.122824\left(  \exp\left(\Omega_{\Lambda}\exp\left(\Omega_{\Lambda}\right)\right) -2.457477z^3  \right)\\\nonumber\\\nonumber
D_6 & =  \tan\left(0.734744+\sin(\sin(\sin(z)))\right)-53.56901z^6 + 0.063339z^5\Omega_{\Lambda} + 28.796105z^5 \\\nonumber &- 0.447099z^4\Omega_{\Lambda}^2 + 0.225933z^4\Omega_{\Lambda} - 52.570121z^4 + 0.701332z^3\Omega_{\Lambda}^3 + 0.425537z^3\Omega_{\Lambda}^2 \\\nonumber &- 1.341193z^3\Omega_{\Lambda}+ 40.109239z^3 + 16.501933z^2\Omega_{\Lambda}^4 - 28.655719z^2\Omega_{\Lambda}^3 + 17.824517z^2\Omega_{\Lambda}^2 \\\nonumber &- 1.666104z^2\Omega_{\Lambda} - 15.679364z^2 + 36.886675z\Omega_{\Lambda}^5 - 110.290456z\Omega_{\Lambda}^4 + 119.350553z\Omega_{\Lambda}^3 \\\nonumber &- 57.503621z\Omega_{\Lambda}^2 + 12.128323z\Omega_{\Lambda} + 3.826671z + 36.544074\Omega_{\Lambda}^6 - 132.936184\Omega_{\Lambda}^5 \\\nonumber &+ 191.252582\Omega_{\Lambda}^4 - 137.162651\Omega_{\Lambda}^3 + 50.699342\Omega_{\Lambda}^2 - 8.947051\Omega_{\Lambda} - 0.393428.
\end{align}
\end{widetext}
The complexity and accuracy scores of the symbolic expressions given above are summarized in table \ref{table:D}. As seen from table \ref{table:D}, the expressions are listed with successively larger complexity from top to bottom in the above.
\newline\newline
The phenomenological symbolic expressions for the Hubble parameter obtained with AI Feynman and PySR are (with the fits corresponding to the Hubble parameter given in units of $1/\rm Gyr$)
\begin{widetext}
\begin{align}\label{eq:H}
	H_1 & = 0.0164896z^2 - 0.136823z\Omega_{\Lambda} + 0.130443z - 0.063049\Omega_{\Lambda}^2 + 0.074300\Omega_{\Lambda} + 0.052666\\\nonumber\\\nonumber
	H_2 & = 0.000898z^4 + 0.002113z^3\Omega_{\Lambda}- 0.007298z^3 - 0.019899z^2\Omega_{\Lambda}^2 + 0.004782z^2\Omega_{\Lambda} + 0.033231z^2 \\\nonumber &- 0.105335z\Omega_{\Lambda}^3 + 0.122373z\Omega_{\Lambda}^2- 0.142459z\Omega_{\Lambda} + 0.108916z - 0.082616\Omega_{\Lambda}^4 + \\\nonumber &0.211217\Omega_{\Lambda}^3 - 0.174108\Omega_{\Lambda}^2 + 0.052916\Omega_{\Lambda} + 0.067133\\\nonumber\\\nonumber
	H_3 & =  \frac{z}{100}\left[    \left\langle  13.607373 + ( \left\lbrace z + \Omega_{\Lambda}\right\rbrace  \cdot \left\lbrace \Omega_{\Lambda} + 0.788210\right\rbrace  )\right\rangle  \cos(0.581035 + \Omega_{\Lambda})   + z\right]  + 0.069529514\\\nonumber\\\nonumber
	H_4 & = 0.127500\left[  z+\cos\left(  \Omega_{\Lambda}\tan^{-1}(z)+1  \right) \right] .
\end{align}
\end{widetext}
As with the expressions fitting the redshift-distance relation, the expressions for the Hubble parameter are ordered with the complexity increasing from top to bottom. The complexity and accuracy of the expressions are summarized in table \ref{table:H}. One may notice that the complexity values for the Hubble data are generally smaller than for the redshift-distance data and also much more closely placed which is another way in which the two data sets complement each other. While six different phenomenological expressions approximating the redshift-distance measure were considered, only four different expressions are used for the Hubble parameter. This is simply because the phenomenological expressions obtained for the Hubble parameter tended to be much less diverse regarding complexity and accuracy than the expressions found for the redshift-distance relation. It was therefore assessed that not much was gained from including more expressions for the Hubble parameter since these would be very similar to the four shown in table \ref{table:H} in terms of complexity and accuracy.
\newline\newline

\begin{figure*}
	\centering
	\subfigure[]{
		\includegraphics[scale = 0.5]{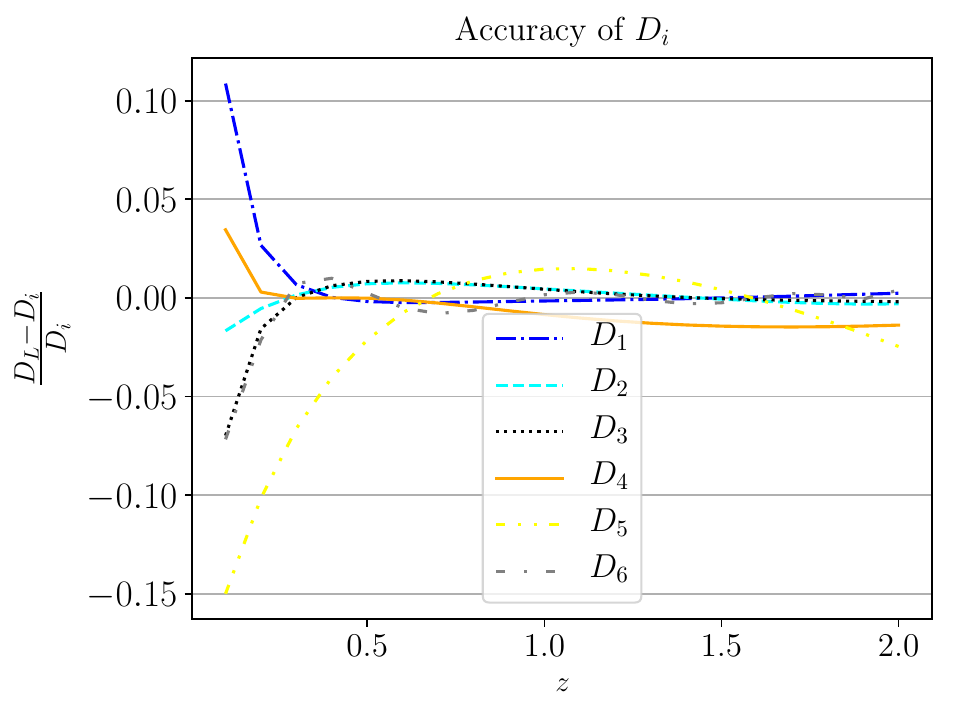}
	}
	\subfigure[]{
		\includegraphics[scale = 0.5]{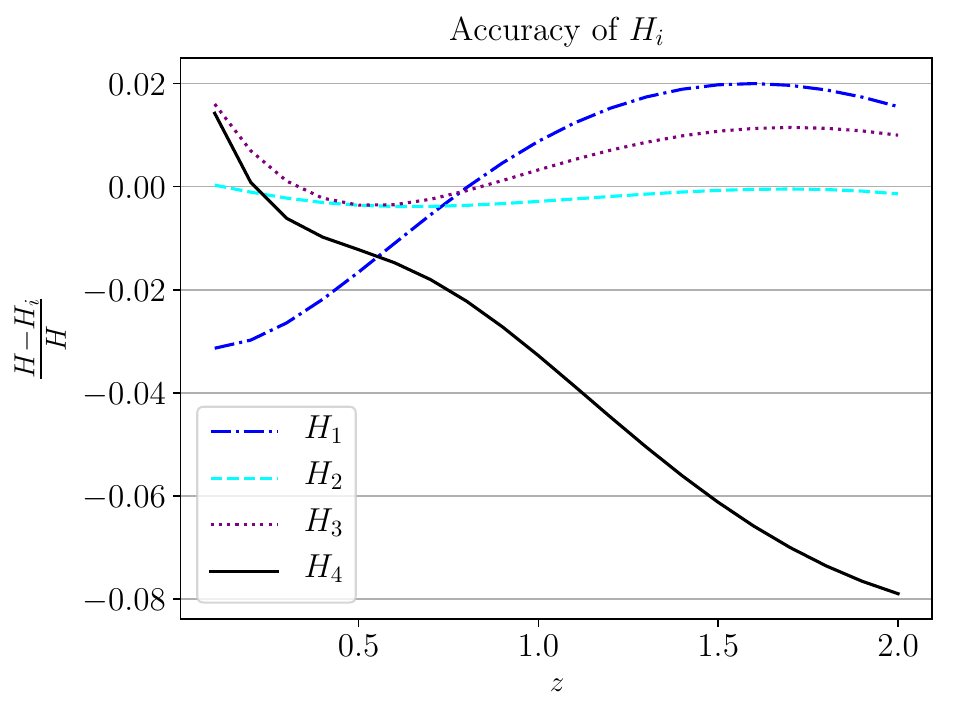}
	}
	\caption{ Accuracy of $H_i$ and $D_i$ for $\Omega_{\Lambda} = 0.7$.}
\label{fig:HDaccuracy}
\end{figure*}

\begin{figure}[tbp]
	\centering 
	\includegraphics[scale = 0.5]{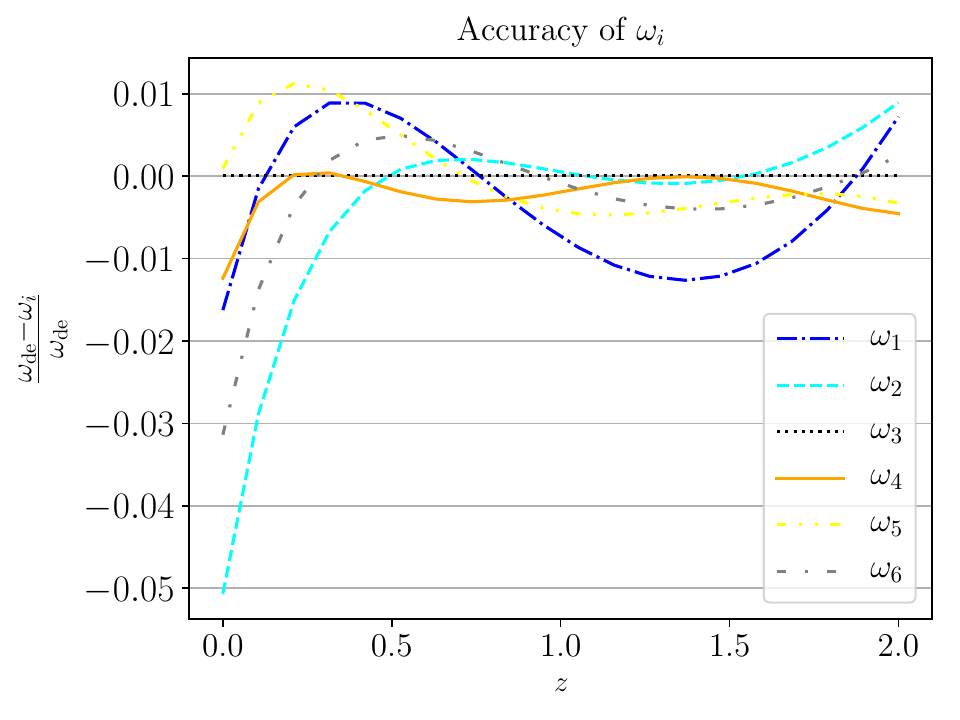}
	\caption{\label{fig:eos_accuracy} Accuracy of $\omega_i$ for $\Omega_{\Lambda} = 0.7$.}
\end{figure}

The phenomenological expressions approximating the equation-of-state parameter in equation \ref{eq:omegaDE} obtained with AI Feynman and PySR are
\begin{widetext}
\begin{align}
	\omega_1 & = -1.010065 + 0.209301z -0.192064\Omega_{\Lambda}\\\nonumber
	\omega_2 &= -2.417910+\ln(z+\pi-\sin\left[ \Omega_{\Lambda}-1\right]  )\\\nonumber
	\omega_3 & = -0.367855\exp((z/(z+1)))/(z+1) ((\Omega_{\Lambda}+1)\exp(1.0/(\Omega_{\Lambda}+1)))\\\nonumber
	\omega_4 &= \ln\left( 0.085759796 \left[   \exp\left( \cos\left\lbrace \sin\left( \Omega_{\Lambda}\right) \right\rbrace \right)  + 0.81087404 + \cos(\cos(\sin(1.067694z) -0.635330)) + z\right] \right) \\\nonumber
	\omega_5 &= -1.820523+1.0/(\tan^{-1}(\exp(\exp(\Omega_{\Lambda}-z))))\\\nonumber
	\omega_6 &  = -1.233155 + \sin(\sin(0.242236(\cos(\cos(\cos(\sin(\Omega_{\Lambda} -0.341135) \cos(\Omega_{\Lambda}))) + \Omega_{\Lambda}) + z))).
\end{align}
\end{widetext}
Note that expression $\omega_3$, found by AI Feynman, is nearly the correct physical expression given in equation \ref{eq:omegaDE}.
\newline\newline
Table \ref{table:omega} shows the accuracy and complexity scores of the six phenomenological expressions for $\omega_{\rm de}$.
\newline\newline
To supplement the accuracy values given in the tables, the accuracies of $D_i$ and $H_i$ are shown for the parameter value $\Omega_{\Lambda}=0.7$ (the value used when computing mock data) in figure \ref{fig:HDaccuracy} along the redshift interval $z\in[0,2]$. Similarly, the accuracies of $\omega_i$ are illustrated in figure \ref{fig:eos_accuracy}. As seen from figure \ref{fig:eos_accuracy}, the phenomenological models for $\omega_{\rm de}$ are all very accurate, with accuracies generally below or around 1\%, with the highest inaccuracy (``achieved'' by $\omega_5$) being around 5\% and reached only at low redshifts.

\begin{table}[tbp]
	\centering
	\begin{tabular}{|l|rccc|}
		\hline
		Model& accuracy.20 & accuracy.all & complexity& number of $\pm$\\
		\hline
		$\omega_1$ & 0.000511 &53.000022 &5 & 2\\
		$\omega_2$ & 0.000800 &73.217119 &2049 & 4\\
		$\omega_3$ & $2.95\cdot 10^{-8}$ &0.001406 &36864 &4\\
		$\omega_4$ & 0.000041 & 3.017021 &$2^{32}+3\approx4\cdot10^{9}$ & 4\\
		$\omega_5$ & 0.000314 &5.050875 &$2^{65536}+1$ & 2\\
		$\omega_6$ & 0.000216 &27.481365 &$\sim 2^{10^{24}}$ & 4\\
		\hline
	\end{tabular}
	\caption{Accuracy scores and complexity of phenomenological symbolic expressions (marked as ``models'' in the table) approximating the equation-of-state parameter for the considered dark energy model. The first accuracy score, ``accuracy.20'' indicates the accuracy obtained when only summing over the 20 mock data points, evaluating at the best-fit value of $\Omega_{\Lambda}$. The other accuracy score, ``accuracy.all'', was obtained as the sum of the least square errors of one million data points computed equidistantly in the intervals $z\in[0,2]$ and $\Omega_{\Lambda}\in[0.1,0.95]$.}
\label{table:omega}
\end{table}

\section{Results}\label{sec:results}
This section presents the results from using Markov chain Monte Carlo (MCMC) sampling to constrain $\Omega_{\Lambda}$ with (mock) Hubble and redshift-distance data using the phenomenological models presented in the previous section as well as by using the derived FLRW expressions.
\newline\indent
The publicly available code emcee\footnote{https://emcee.readthedocs.io/en/stable/} \cite{emcee} is used for MCMC sampling on the mock data.
\newline\newline
Results have been confirmed by rerunning MCMC codes and recomputing mock data with different random seeds. The results from using different random seeds are overall in agreement so only one set of results is shown in the following.

\subsection{Parameter inference using the redshift-distance relation}
This subsection serves to present the parameter constraints obtained by comparing mock redshift-distance data with the phenomenological models for the redshift-distance relation.
\begin{figure}
	\centering
	\includegraphics[scale = 0.5]{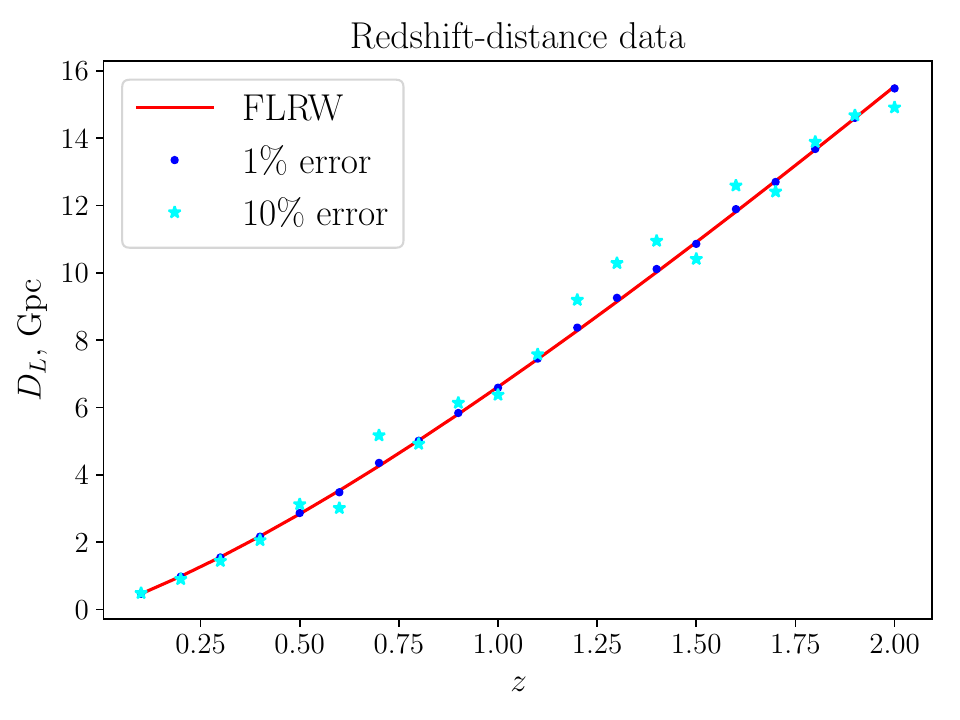}
	\caption{Redshift-distance relation data (dots and stars) together with a curve representing the exact redshift-distance relation for the considered $\Lambda$CDM model.}
	\label{fig:DLdata}
\end{figure}
The mock redshift-distance data was generated as 20 equidistant redshift values $z = 0.1,0.2...,1.9,2$ with corresponding luminosity distance. Two data sets were generated, with errors drawn from a Gaussian distribution corresponding to standard deviations of $1\%$ and $10\%$, respectively. The model used for generating the data was a flat $\Lambda$CDM model with $H_0 = 70$km/s/Mpc and $\Omega_{\Lambda} = 0.7$. The mock data sets are shown in figure \ref{fig:DLdata} together with curves representing the exact (precise) redshift-distance relation.
\newline\newline
Table \ref{table:D_constraints} shows the constraints obtained on $\Omega_{\Lambda}$ using the different expressions for the redshift-distance relation together with emcee. Note that when computing the MCMC chains a small number (1-2 hand fulls out of thousands of points) of $D_3$ and $D_4$ evaluations lead to {\em Nan} due to evaluations of $\sin^{-1}(x)$ with $|x|>1$. Since only very few points lead to errors, this is not expected to have any significant influence on the results.
\newline\newline
\begin{table}[tbp]
	\centering
	\begin{tabular}{|l|rc|}
		\hline
		Model& $\Omega_{\Lambda}$, 1\% error & $\Omega_{\Lambda}$, 10 \% error \\
		\hline
		FLRW & $0.700546^{+0.000299}_{-0.000302}$  & $0.705392^{+0.002962}_{-0.002954}$ \\
		$D_1$ & $0.703242^{+0.000300}_{-0.000294}$ & $0.708184^{+0.002943}_{-0.002852}$ \\
		$D_2$ & $0.698034^{+0.000288}_{-0.000287}$ & $0.702607^{+0.002859}_{-0.002811}$ \\
		$D_3$ & $0.700137^{+0.000284}_{-0.000287}$ & $0.704552^{+0.002828}_{-0.002789}$ \\
		$D_4$ &$0.683405^{+0.000310}_{-0.000311}$  & $0.688576^{+0.003098}_{-0.003091}$ \\
		$D_5$ & $0.681860^{+0.000318}_{-0.000319}$ & $0.686945^{+0.003155}_{-0.003155}$\\
		$D_6$ & $0.751960^{+0.000288}_{-0.000291}$  & $0.756677^{+0.002831}_{-0.002837}$ \\
		\hline
	\end{tabular}
	\caption{Constraint on $\Omega_{\Lambda}$ obtained with emcee from the mock Hubble data sets combined with the different phenomenological expressions as well as from the derived FLRW expression for $\omega_{\rm de}$. The mean was computed as the 50th percentile of the data while the errors correspond to the 15.9th and 84.1st percentiles which corresponds to the ordinary 1 sigma standard deviation if the MCMC data is distributed according to a Gaussian.}\label{table:D_constraints}
\end{table}
From table \ref{table:D_constraints} it is seen that for the very precise data with 1\% error, the correct $\Omega_{\Lambda}$ value of $0.7$ generally doesn't lie inside the 68\% confidence interval, although it is close to doing so. The correct value is contained only within the 68\% confidence interval of $D_3$, while the mean values and confidence intervals obtained using $D_4, D_5$ and $D_6$ are quite far from being correct. Referring back to figure \ref{fig:HDaccuracy} and table \ref{table:D}, this may be a result of models $D_4, D_5$ and $D_6$ being less accurate approximations of $D_L$ than the other phenomenological models, but it could also be a result of the large complexities of these models. The constraints obtained using $D_6$ are much less accurate than those obtained from $D_4$ and $D_5$ which are at the same time much less accurate approximations of $D_L$ than $D_6$ is. This indicates that the higher complexity of $D_6$ is partly the reason for the poor resulting constraints on $\Omega_{\Lambda}$ with $D_6$; $D_6$ has a much larger complexity and number of terms than $D_4$ and $D_5$. This suggestion should be considered with caution; as discussed earlier, the MCMC algorithm utilized here does not obviously contain a part that could lead to significantly different results depending on function complexity.

\begin{figure*}
	\centering
	\subfigure[]{
		\includegraphics[scale = 0.5]{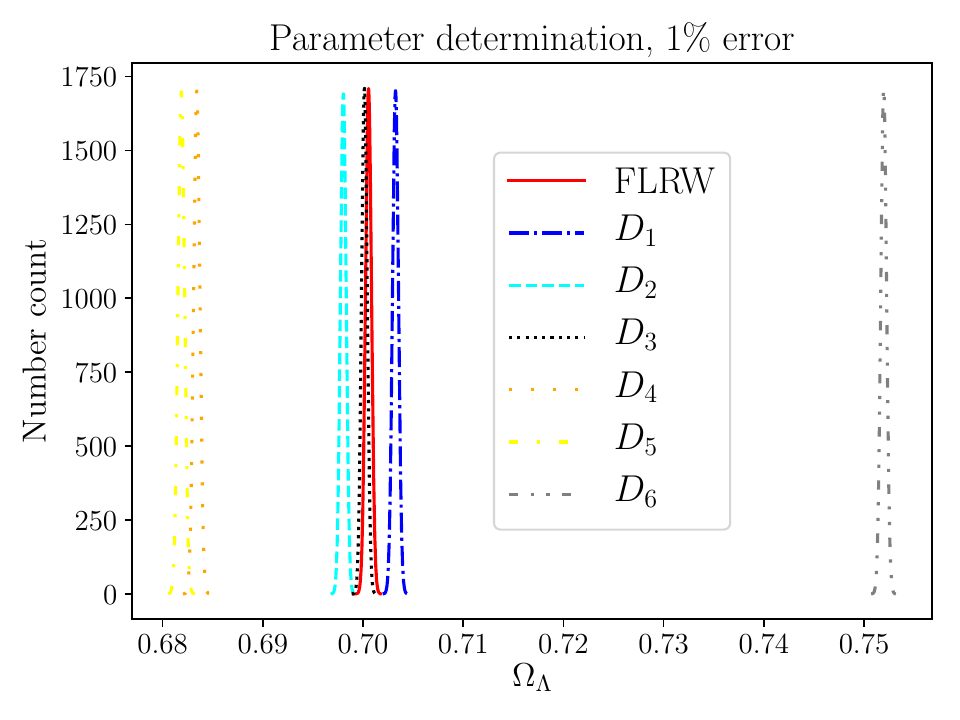}
	}
	\subfigure[]{
		\includegraphics[scale = 0.5]{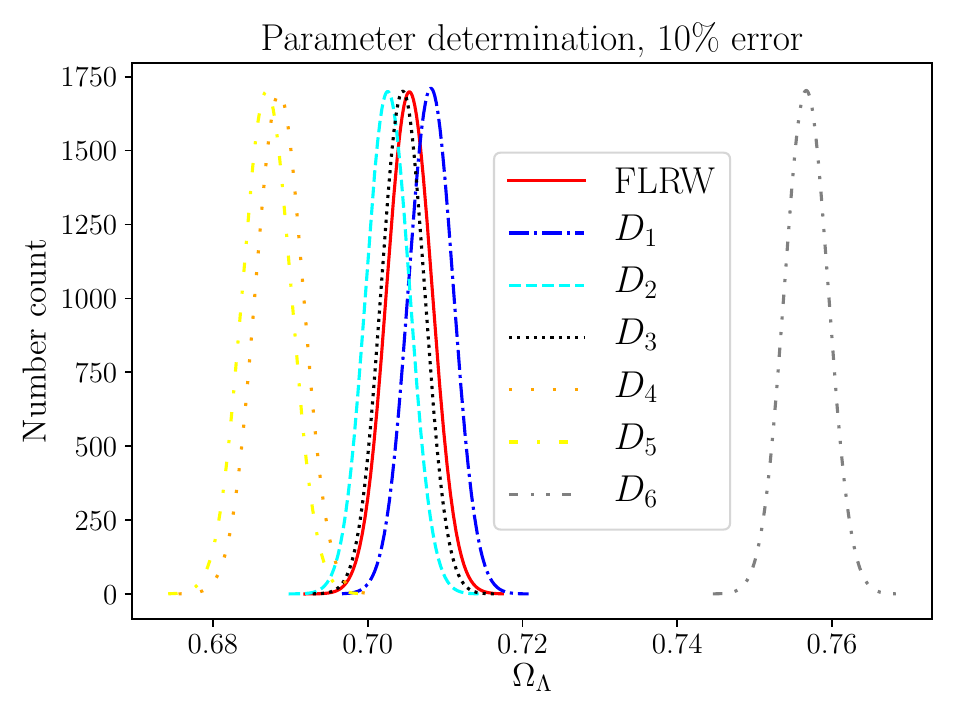}
	}
	\caption{Number count distributions of $\Omega_{\Lambda}$ in MCMC chains obtained with emcee obtained using the different phenomenological expressions as well as the derived FLRW relation for the redshift-distance relation.}
\label{fig:Dgaussian}
\end{figure*}

\subsection{Parameter inference using the Hubble parameter}
\begin{figure}
	\centering
	\includegraphics[scale = 0.5]{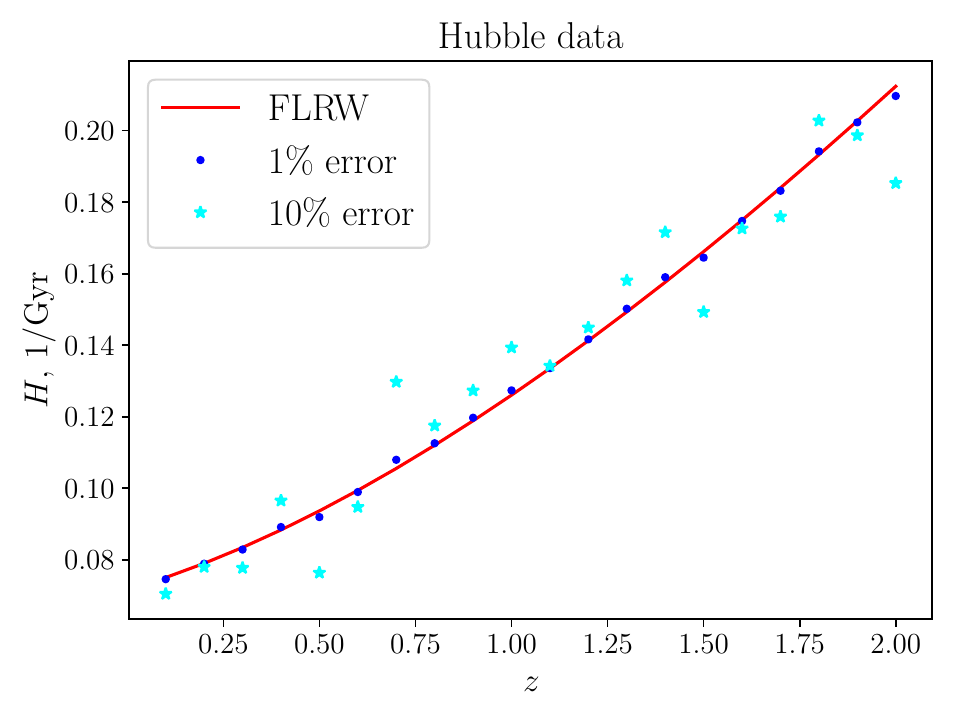}
	\caption{Hubble parameter data (dots and stars) together with a curve representing the exact Hubble-redshift relation for the considered $\Lambda$CDM model.}
	\label{fig:Hdata}
\end{figure}
\begin{figure*}[tbp]
	\centering 
	\subfigure[]{
	\includegraphics[scale=0.5]{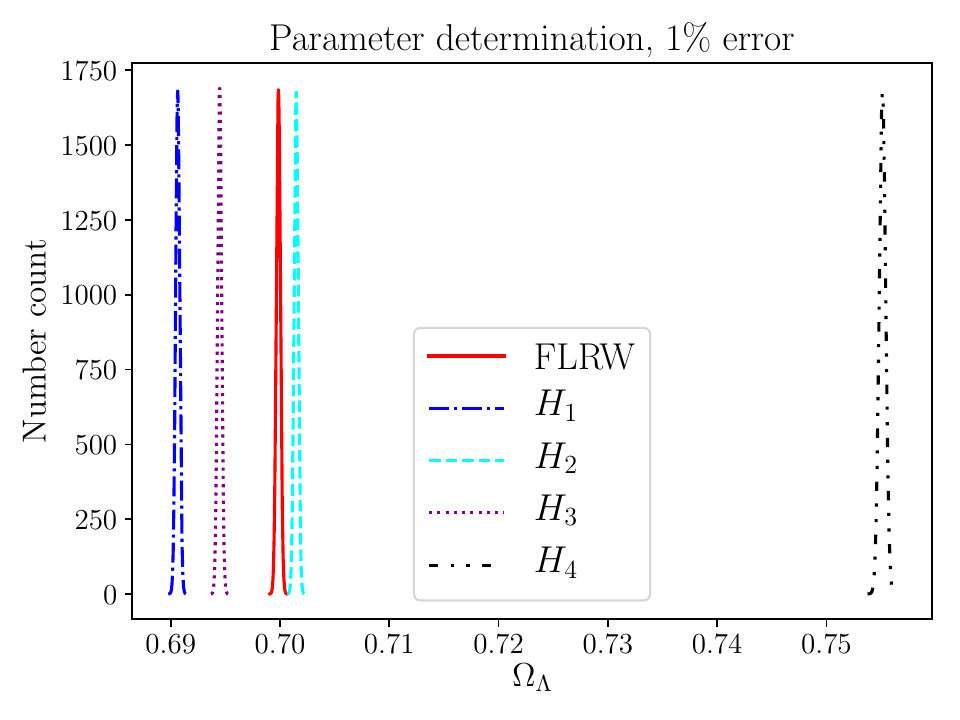}
}
	\subfigure[]{
	\includegraphics[scale=0.5]{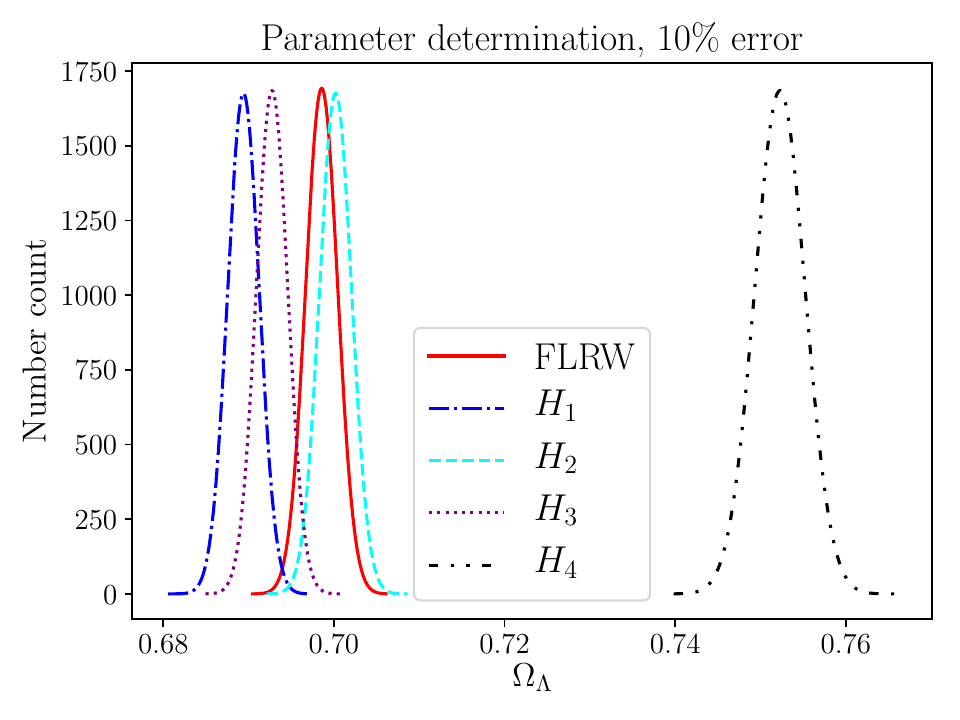}
}
	\caption{\label{fig:Hgaussian} Number count distributions of $\Omega_{\Lambda}$ in MCMC chains obtained with emcee obtained using the different phenomenological expressions as well as the derived FLRW relation for the Hubble parameter.}
\end{figure*}

As with the redshift distance data, the Hubble data is generated at 20 equidistant redshift values $z = 0.1,0.2...,1.9,2$, using a flat $\Lambda$CDM model with $H_0 = 70$km/s/Mpc and $\Omega_{\Lambda} = 0.7$, with errors drawn from Gaussian distributions corresponding to a standard deviation of $1\%$ and $10\%$, respectively. The mock data sets are shown in figure \ref{fig:Hdata} together with a curve representing the precise relation.
\newline\newline
Table \ref{table:H_constraints} shows the constraints obtained for $\Omega_{\Lambda}$ in terms of mean values and standard deviation. As seen, model $H_4$ yields a clearly incorrect result: The mean is almost 10 \% off from the correct value which is far from being contained within the even the 5$\sigma$ confidence interval. Closer inspection shows that several of the other phenomenological models also yield results which do not contain the correct value within their $5\sigma$ confidence intervals. In fact, $H_2$ is the only phenomenological model which contains the correct value within its 68\% confidence interval, and only for the data with 10\% error. The FLRW results both agree with the correct value within 1 standard deviation.
\newline\indent
Taking the results at face value and comparing with the complexities and accuracies in table \ref{table:H}, the conclusion is that a high accuracy is more important than a low complexity since the phenomenological models are increasingly closer to being correct when their accuracy is increased while a similar statement cannot be made for the complexity. This is most easily realized by comparing table \ref{table:H} with figure \ref{fig:Hgaussian} which shows the distributions of $\Omega_{\Lambda}$ in the MCMC chains obtained with emcee. In addition, there is a striking difference between the accuracy of $H_4$ and the other models while only a modest difference in the complexities. Specifically, $H_4$ has a mean square error roughly between 30 and $10^4$ times larger than the other models while the complexity is only roughly between 2 and 20  times larger than those of the other models. Since $H_4$ leads to strikingly poorer parameter constraints, this suggests that the complexity is less important than the accuracy of the phenomenological models. It is also noteworthy that $H_2$ has a complexity more than 4 times as large as $H_1$ (and has the highest number of $\pm$ of all $H_i$) but is also significantly more accurate. The constraints obtained using $H_2$ are closer to the true value than those obtained with $H_1$.

\begin{table}[tbp]
	\centering
	\begin{tabular}{|l|rc|}
		\hline
		Model& $\Omega_{\Lambda}$, 1\% error & $\Omega_{\Lambda}$, 10 \% error \\
		\hline
		FLRW & $0.69986^{+0.00019}_{-0.00019}$ &$0.69857^{+0.00190}_{-0.00188}$ \\
		$H_1$ & $0.69064^{+0.00018}_{-0.00019}$ & $0.68939^{+0.00185}_{-0.00186}$  \\
		$H_2$ & $0.70148^{+0.00019}_{-0.00019}$ &$0.70019^{+0.00189}_{-0.00187}$ \\
		$H_3$ & $0.69448^{+0.00019}_{-0.00019}$ & $0.69322^{+0.00184}_{-0.00186}$\\
		$H_4$ & $0.75513^{+0.00030}_{-0.00030}$ &$0.75227^{+0.00297}_{-0.00296}$\\
		\hline
	\end{tabular}
	\caption{Constraint on $\Omega_{\Lambda}$ obtained with emcee from the mock Hubble data sets combined with the different phenomenological expressions as well as from the derived FLRW expression. The mean was computed as the 50th percentile of the data while the errors correspond to the 15.9th and 84.1st percentiles which corresponds to the ordinary 1 sigma standard deviation if the MCMC data is distributed according to a Gaussian.}\label{table:H_constraints}
\end{table}

\subsection{Parameter inference using phenomenological expressions for $\omega_{\rm de}$}
In this last subsection, constraints on $\Omega_{\Lambda}$ ($=\Omega_{\rm de,0}$) are obtained by considering mock data generated for a flat matter+dark energy model with $H_0 = 70$km/s/Mpc, $\Omega_{\Lambda} = 0.7$ and $\omega_{\rm de}$ as given by equation \ref{eq:omegaDE}. The mock data was generated as Hubble parameter and redshift-distance data with the same intervals and errors as in the previous sections. In this section, however, the phenomenological models represent $\omega_{\rm de}$ and the Hubble and redshift-distance predictions will be computed using these phenomenological models of $\omega_{\rm de}$ together with equations \ref{eq:Friedmann} and \ref{eq:DL}.
\newline\newline
The two mock data sets were studied independently. The constraints on $\Omega_{\Lambda}$ obtained with the Hubble data (using emcee) are shown in table \ref{table:eos_Hconstraints} while those obtained with redshift-distance data are shown in table \ref{table:eos_DLconstraints}. As seen, the results obtained with the two data sets are very similar. The constraints generally contain the correct value of $\Omega_{\Lambda}$ within a few standard deviations. The most notable result is that the constraints obtained by using $\omega_2$ with the precise mock data (with errors of $1\%$) are clearly worse than those obtained with the other models. Looking at figure \ref{fig:eos_accuracy} it is seen that this model also has a clearly lower accuracy than the other models for the highest and lowest parts of the redshift interval. It is also notable that the models with highest complexity yield results as good as or better than those obtained with the less complex models even though their accuracies are similar. Indeed, model $\omega_5$ which has a very high complexity just barely yields a more accurate mean value of $\Omega_{\Lambda}$ than the correct expression for $\omega_{\rm de}$ does.
\newline\indent
Note lastly that the constraints obtained using $\omega_3$ are almost identical to those obtained using the correct $\omega_{\rm de}$. This is as expected since $\omega_3$ is nearly identical to the true expression for $\omega_{\rm de}$ and thus very accurately fits the true expression.
\begin{table}[tbp]
	\centering
	\begin{tabular}{|l|rc|}
		\hline
		Model& $\Omega_{\Lambda}$, 1\% error & $\Omega_{\Lambda}$, 10 \% error \\
		\hline
		FLRW & $0.699707^{+0.000168}_{-0.000166}$  & $0.697056^{+0.001695}_{-0.001655}$  \\
		$\omega_1$ & $0.699626^{+0.000170}_{-0.000169}$ &  $0.696958^{+0.001682}_{-0.001686}$  \\
		$\omega_2$& $0.698837^{+0.000168}_{-0.000169}$  &  $0.696190^{+0.001699}_{- 0.001677}$\\
		$\omega_3$ & $0.699713^{+0.000167}_{-0.000167}$  & $0.697070^{+0.001679}_{-0.001676}$ \\
		$\omega_4$ & $0.699484^{+0.000168}_{-0.000168}$  &  $0.696840^{+0.001689}_{-0.001680}$\\
		$\omega_5$ & $0.700066^{+0.000167}_{-0.000169}$  &  $0.697403^{+0.001702}_{-0.001688}$\\
		$\omega_6$ & $0.699391^{+0.000167}_{-0.000167}$  &  $0.696751^{+0.001672}_{-0.001678}$\\
		\hline
	\end{tabular}
	\caption{Constraint on $\Omega_{\Lambda}$ obtained with emcee from the mock Hubble data sets combined with the different phenomenological expressions as well as from the derived FLRW expression. The mean was computed as the 50th percentile of the data while the errors correspond to the 15.9th and 84.1st percentiles which corresponds to the ordinary 1 sigma standard deviation if the MCMC data is distributed according to a Gaussian.}\label{table:eos_Hconstraints}
\end{table}
\begin{table}[tbp]
	\centering
	\begin{tabular}{|l|rc|}
		\hline
		Model& $\Omega_{\Lambda}$, 1\% error & $\Omega_{\Lambda}$, 10 \% error \\
		\hline
		FLRW & $0.700486^{+0.000270}_{-0.000270}$  & $0.704846^{+0.002716}_{- 0.002671}$\\
		$\omega_1$ &  $0.700656^{+0.000274}_{-0.000273}$ &  $0.705105^{+ 0.002736}_{-0.002662}$ \\
		$\omega_2$& $0.696789^{+0.000271}_{-0.000273}$  & $0.701064^{+ 0.002675}_{-0.002687}$ \\
		$\omega_3$ & $0.700500^{+0.000272}_{-0.000273}$ & $0.704868^{-0.002686}_{-0.002641}$ \\
		$\omega_4$ & $0.699890^{+0.000272}_{-0.000267}$  & $0.704232^{+0.002656}_{-0.002654}$ \\
		$\omega_5$ &  $0.701981^{+0.000278}_{-0.000274}$& $0.706393^{+0.002739}_{-0.002700}$ \\
		$\omega_6$ & $0.699004^{+0.000275}_{-0.000273}$ & $0.703348^{+0.002685}_{-0.002641}$\\
		\hline
	\end{tabular}
	\caption{Constraint on $\Omega_{\Lambda}$ obtained with emcee from the mock redshift-distance data sets combined with the different phenomenological expressions as well as from the derived FLRW expression. The mean was computed as the 50th percentile of the data while the errors correspond to the 15.9th and 84.1st percentiles which corresponds to the ordinary 1 sigma standard deviation if the MCMC data is distributed according to a Gaussian.}\label{table:eos_DLconstraints}
\end{table}

\begin{figure*}[tbp]
	\centering 
	\includegraphics[scale=0.5]{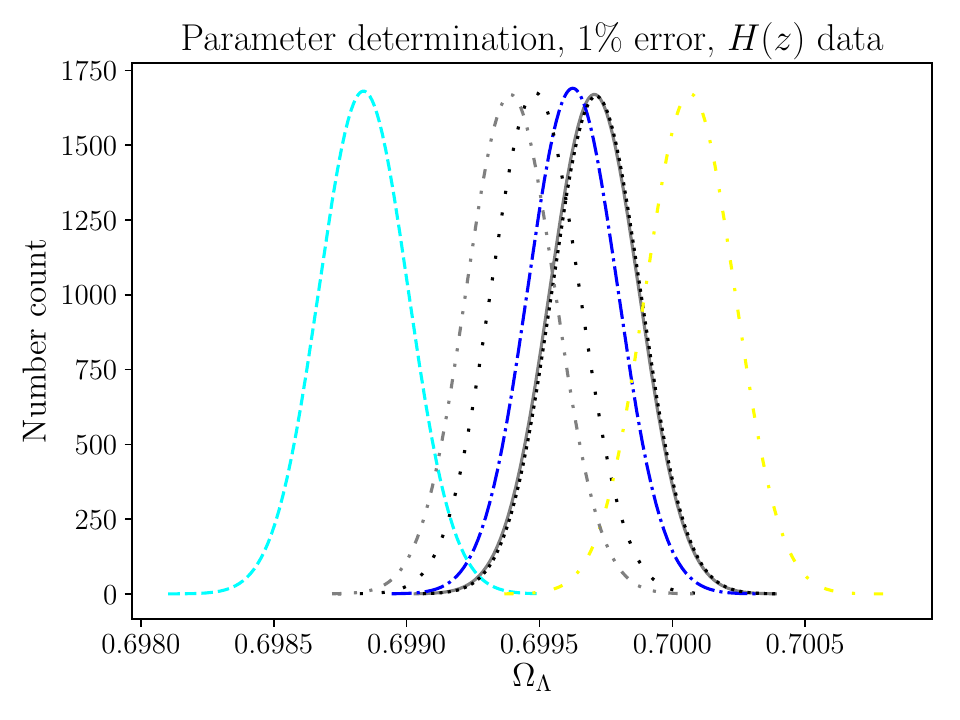}
	\includegraphics[scale=0.5]{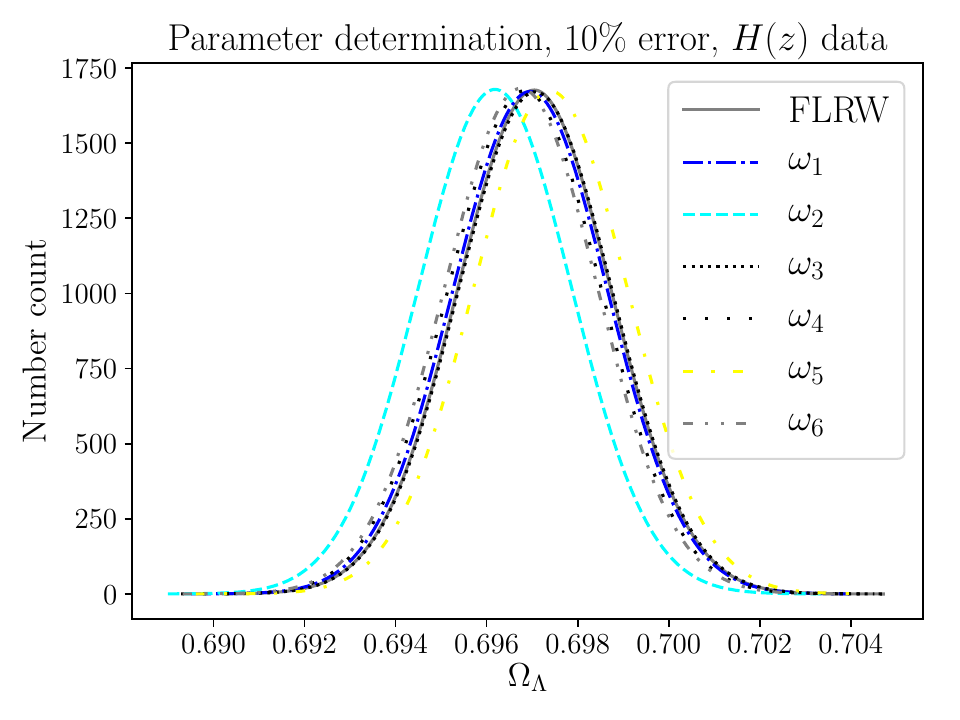}
	\par
	\includegraphics[scale=0.5]{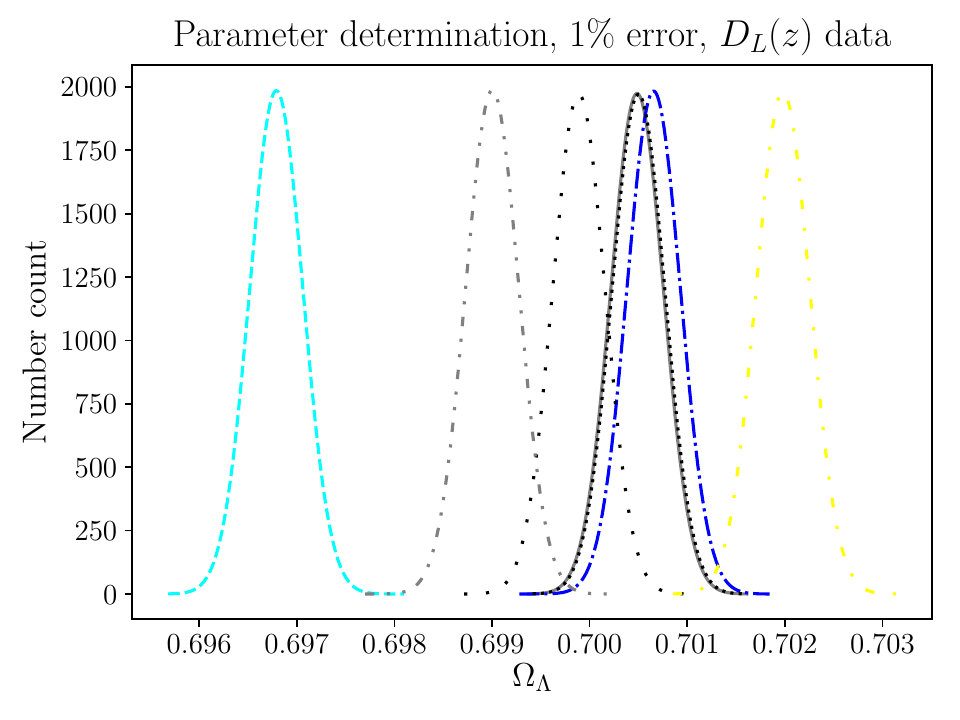}
	\includegraphics[scale=0.5]{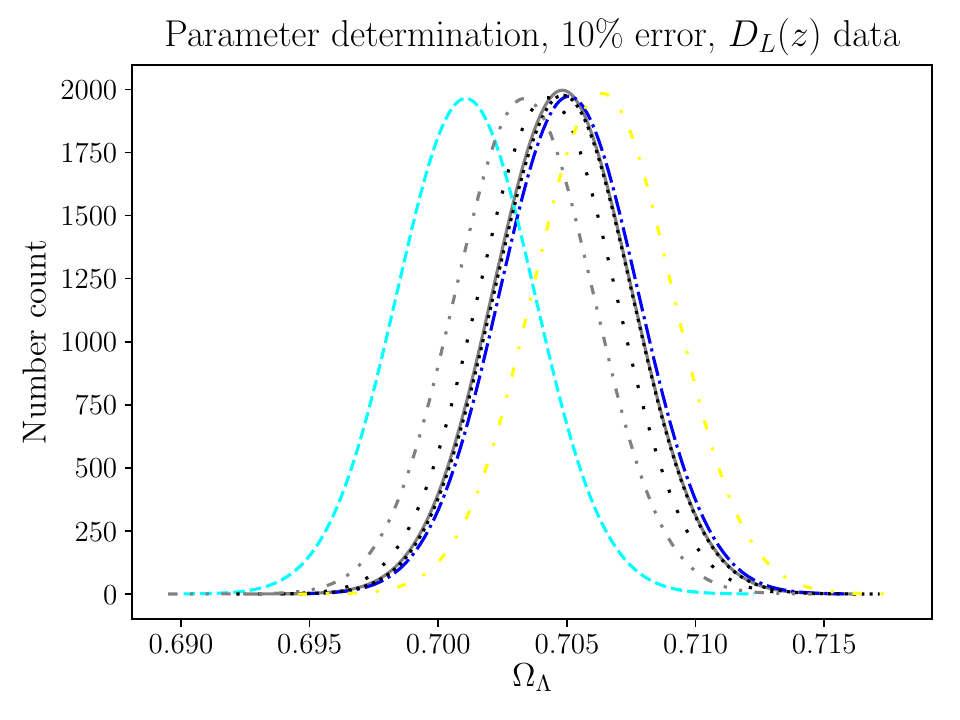}
	\caption{\label{fig:Heosgaussian} Number count distributions of $\Omega_{\Lambda}$ in MCMC chains obtained with emcee obtained using the different phenomenological expressions as well as the correct expression for $\omega_{\rm de}$ combined with mock Hubble (top figures) and redshift-distance (bottom figures) data.}
\end{figure*}

\section{Summary and conclusions}\label{sec:summary}
Constraints on the cosmological parameter $\Omega_{\Lambda}$ were obtained by considering mock Hubble and redshift-distance data with two levels of data accuracy (1\% and 10\%). The constraints were obtained using different phenomenological expressions approximating the observables and the equation-of-state parameter of dark energy as well as by using the physically motivated, derived relations. An array of phenomenological models with different levels of complexity and accuracy were considered. By comparing the results obtained using the different models and data sets, it appears from the parameter constraints based on phenomenological redshift-distance relations that the complexity of the phenomenological models may become important if the complexity is very high. In general though, it is the accuracy which mainly affects the models' abilities to constrain $\Omega_{\Lambda}$ correctly, and the constraints obtained based on Hubble data and the phenomenological equation-of-state parameter of dark energy do not indicate that the complexity of the phenomenological models affect the results. The rigidity of this conclusion must not be considered too firm though. For instance, it is possible that the significance of the complexity of a phenomenological model depends on how much of the trend of a relation is expressed through the more complex terms. This could e.g. be assessed by computing the accuracy of the individual terms of a phenomenological expression in terms of reproducing the entire true relation. In addition, the literature contains several proposals as to how to best encapsulate the complexity of an expression (see e.g. \cite{complexity}). It cannot be excluded that another measure of complexity than the ones used here would reveal a stronger correlation between incorrectness of parameter constraints and high complexity (but note that such a correlation is missing both when using the complexity as defined in equation \ref{eq:complexity} as well as when using the number of terms as the measure of complexity). However, with the complexity measure used here, the overall conclusion which can be drawn from the current study is that the complexity of a phenomenological expression does not seem to have significant impact on parameter constraints unless the complexity is very high. When considering the MCMC algorithms overall, the most surprising part of this conclusion is that the results indeed may be affected by high complexity; there does not seem to be an obvious reason why MCMC algorithms would lead to statistically significantly different results when using different phenomenological expressions with identical accuracy.
\newline\indent
The accuracy of the expressions generally seems to be much more important and the different constraints obtained by using different phenomenological models do not generally agree regarding their mean value and sometimes up to several standard deviations. This indicates that it would be good practice to consider constraints using more than a single phenomenological model when using these, in order to assess the robustness of ones findings. Naturally, this becomes more important when the data being used for obtaining the parameter constraints becomes increasingly precise. An example where this could be particularly important is when constraining the dark energy equation-of-state where non-vacuum models are usually phenomenological in which case there is no reason to expect that the true/physical equation-of-state parameter will turn out to be exactly as modeled. A possible method could be to fit the ``theoretical'' equation-of-state parameter of interest using symbolic regression (much as done here) and repeat computations with these to assess robustness of findings. Similarly, when using cosmographic expansions, it would be beneficial to use expansions at different order or using different parameterizations to cross-validate ones results.
\newline\newline
It is lastly noted that while an aim here was to understand if the complexity of a phenomenological expression has significant effect on parameter constraints, there are other, entirely different, reasons to seek low complexity, namely computation time. Computing the MCMC chains required much less computation time when using the phenomenological expressions for $D_L$ than when using $D_L$ itself, since the latter depends on an integral which is evaluated numerically. When comparing the individual phenomenological models, the expressions with the fewest terms were fastest which justifies why complexity definitions used in some computer algorithms are solely based on number of terms. For instance taking $10^4$ steps with emcee using model $D_1$ took 25 seconds while the same only required 11 seconds of computing time for model $D_2$. $D_1$ has 20 $\pm$ signs and a complexity of 417 while $D_2$ has 4 $\pm$ signs and a complexity of 658.

\section{Acknowledgments}
The author is funded by VILLUM FONDEN, grant VIL53032. Part of the numerical work done for this project was performed using the UCloud interactive HPC system managed by the eScience Center at the University of Southern Denmark.


\appendix
\section{Example of complexity computation}\label{app:tree_complexity}
\begin{figure*}
	\centering
		\includegraphics[scale = 0.63]{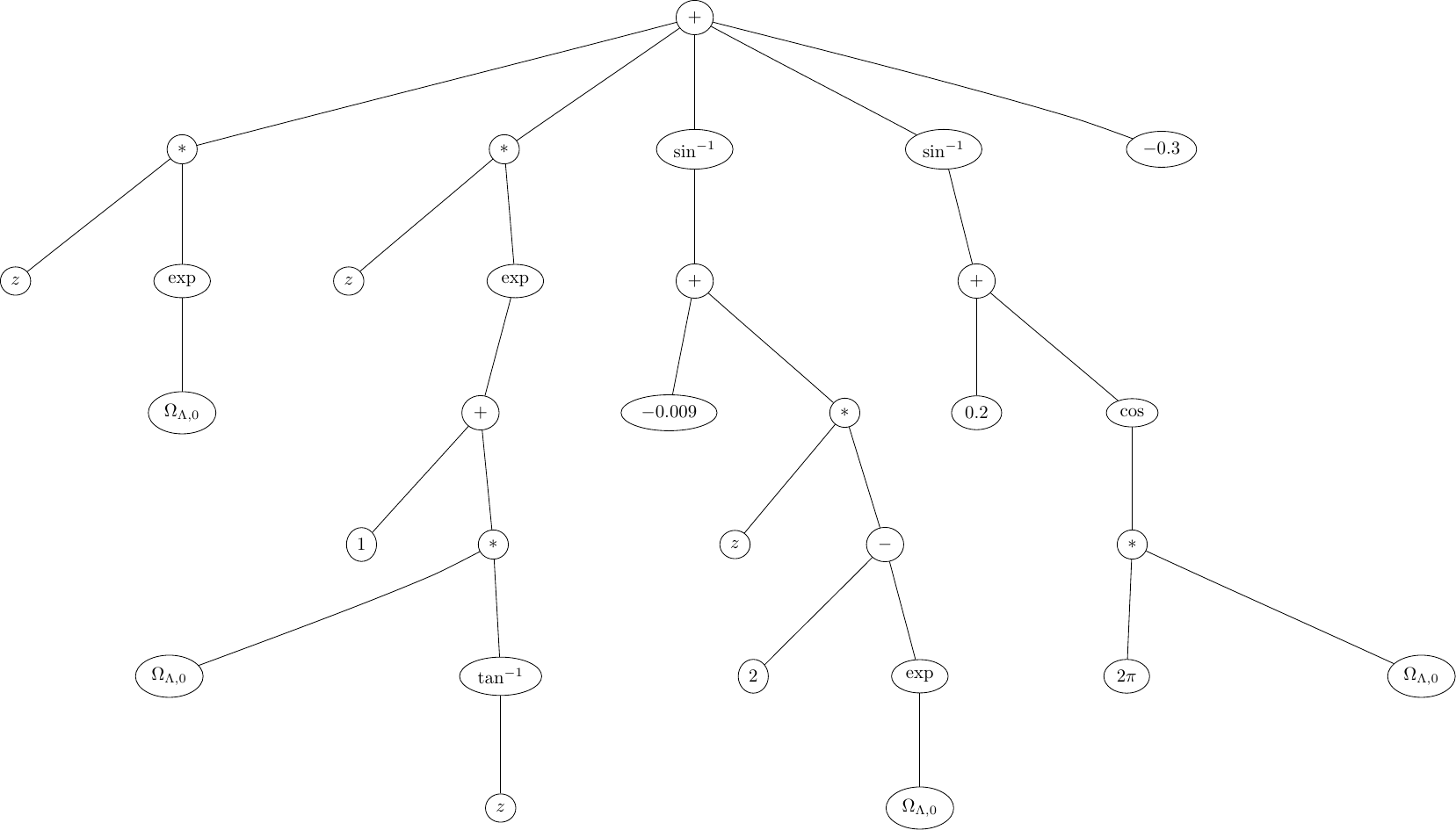}
	\caption{Symbolic expression tree for $D_4$.}
	\label{fig:tree}
\end{figure*}

This appendix serves to demonstrate how the complexity of a symbolic expression is computed using the formula given in equation \ref{eq:complexity}.
\\\\
Figure \ref{fig:tree} shows the symbolic expression tree for the expression $D_4$. To compute the complexity of $D_4$ we must start at the bottom node of each branch and see if the node contains a variable ($\Omega_{\Lambda}$ or $z$), or a constant. This gives us the value of complexity$(c=1)$ which we can iteratively use to ``climb'' the tree until we reach the top. Starting at the left branch, we see that there are two end-nodes: One with a $z$ and one with an $\Omega_{\Lambda}$. Each represent a variable and therefore contribute with a complexity value of 2 (each). The $\Omega_{\Lambda}$ node is connected to an ``$\exp$'' node which has a complexity we compute as $2^2=4$. The next node has a multiplication sign so the complexity of the total complexity from this branch is obtained by multiplying the complexities of the two sub-branches, i.e. as $2\cdot 4 = 8$. Proceeding similarly with the remaining branches, they are seen to have complexities (from left to right)  1024, 2048, 32 and 1. The total complexity of the tree is thus complexity = $8+1024+2048+31+1=3113$. Other examples of symbolic expression trees and corresponding complexities can be found in \cite{complexity}.

\end{document}